\documentclass[12pt]{iopart}
\usepackage{graphicx}

\newcommand{\ud}{\mathrm{d}}

\begin{document}

\title{Heavy quark(onium) at LHC: the statistical hadronization case}

\author{A Andronic$^1$, P Braun-Munzinger$^{1,2,3,4}$, K Redlich$^{1,5,3}$, 
J Stachel$^6$}
\address{$^1$~GSI Helmholtzzentrum f\"ur Schwerionenforschung,
Darmstadt, Germany, 
$^2$~Technical University, Darmstadt, Germany
$^3$~ExtreMe Matter Institute EMMI, GSI, Darmstadt, Germany
$^4$~Frankfurt Institute for Advanced Studies, J.W. Goethe University,
Frankfurt, Germany,
$^5$~Institute of Theoretical Physics, University of Wroc\l aw, Poland,
$^6$~Physikalisches Institut, University of Heidelberg, Germany}


\begin{abstract}
We discuss the production of charmonium in nuclear collisions within the 
framework of the statistical hadronization model. 
We demonstrate that the model reproduces very well the availble data
at RHIC. We provide predictions for the LHC energy where, dependently 
on the charm production cross section, a dramatically different behaviour 
of  charmonium production as a function of centrality might be expected.
We discuss also the case in elementary collisions, where clearly the 
statistical model does not reproduce the measurements.
\end{abstract}



\section{Introduction}
Charmonium production is considered, since the original proposal more than 20
years ago about its suppression in a Quark-Gluon Plasma (QGP)
\cite{satz}, an important probe to determine the degree of deconfinement
reached in the fireball produced in ultra-relativistic nucleus-nucleus
collisions. 
In the original scenario of J/$\psi$ suppression via Debye screening
\cite{satz} it is assumed that the charmonia are rapidly formed in initial
hard collisions but are subsequently destroyed in the QGP (see recent summaries
in \cite{satz_kluberg,rapp_rev}).

In recent publications \cite{aa2}  we have demonstrated that the data on 
J/$\psi$ and $\psi'$ production in nucleus-nucleus 
collisions at the SPS ($\sqrt{s_{NN}} \approx 17$ GeV) and RHIC 
($\sqrt{s_{NN}}$=200 GeV) energies can be well described within the statistical 
hadronization model proposed in \cite{pbm1} and have provided predictions for 
the LHC energy ($\sqrt{s_{NN}}$=5.5 TeV) and for energies close to threshold 
($\sqrt{s_{NN}} \approx 6$ GeV).
Because the mass of heavy quarks exceeds the transition
temperature of the QCD phase transition by a factor of almost 8, heavy
flavor hadron production cannot be described in a purely thermal approach. 
It was, however, realized in \cite{pbm1} that charmonium and charmed hadron
production can be well described by assuming that all charm quarks are
produced in initial, hard collisions while charmed hadron and charmonium
production takes place exclusively at the phase boundary with statistical 
weights calculated in a thermal approach. For a recent review of this 
statistical hadronization approach see \cite{pbm_js_lb}. 
An important element is thermal equilibration, at least near the critical 
temperature, $T_c$, which we believe can be achieved efficiently for charm 
only in the QGP. 

An important question is whether the statistical behavior is a unique feature 
of high energy nucleus-nucleus collisions or whether it is also encountered 
in elementary collisions, where finite size effects will obscure
a possible phase transition.
In early analyses (see ref.~\cite{becattini97}), it is indeed argued that 
hadron production in e$^+$e$^-$ and pp is thermal in nature. 
Furthermore, such analyses of hadron multiplicities (for recent results 
see \cite{aa08_ee,becattini08,kraus08} and refs. therein) yield also 
temperature values in the range of 160-170 MeV.

To establish the uniqueness of the J/$\psi$ probe for the diagnosis of 
a QGP in nucleus-nucleus collisions  it is important to understand 
whether similar thermal features as observed in nucleus-nucleus collisions
are also at work in heavy-flavor hadron production in elementary collisions.
We have recently demonstrated \cite{aa09} that there are crucial differences
in elementary collisions compared to nucleus-nucleus.

\section{The statistical model in elementary collisions}

For the study of hadron production in e$^+$e$^-$ collisions we employ the 
canonical statistical model described in \cite{aa08_ee,aa09} (see also 
\cite{becattini08,fb09}).
For the present study, we perform calculations for two cases: 
i) a 2-jet initial state which carries the quantum 
numbers of the 5 flavors, with the relative abundance of the five flavors 
in one jet and corresponding antiflavor in the other jet
taken form the measurements at the $Z^0$ resonance quoted in \cite{pdg}.
These relative abundances (17.6\% for $u\bar{u}$ and $c\bar{c}$ and 21.6\% for 
$d\bar{d}$, $s\bar{s}$ and $b\bar{b}$) are thus external input values, 
unrelated with the thermal model.
ii) a purely thermal ansatz , i.e. a 2-jet initial state characterized by 
vanishing quantum numbers in each jet.

\begin{figure}[htb]
\begin{tabular}{cc} \begin{minipage}{.46\textwidth}
\centering\includegraphics[width=1.2\textwidth]{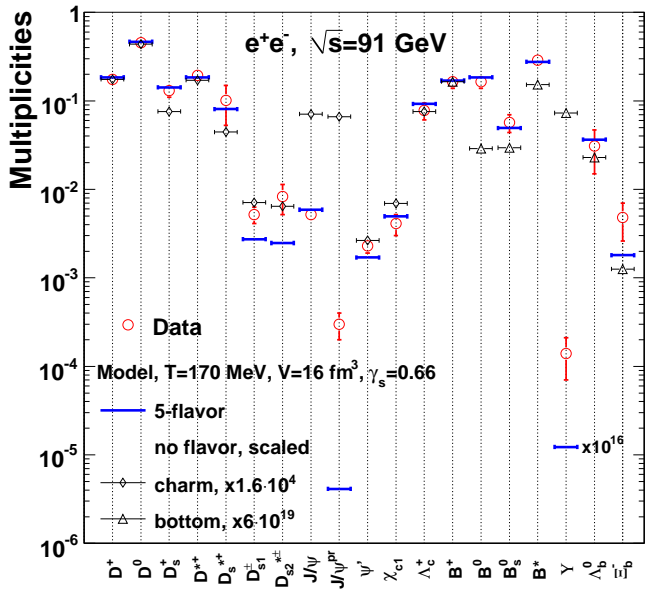}
\end{minipage}  & \begin{minipage}{.49\textwidth}
\caption{Multiplicities of hadrons with charm and bottom quarks in  e$^+$e$^-$ 
collisions compared to the thermal model calculations for two cases: 
i) the 5-flavor jet scheme (thick lines) and ii) no (net) flavor jet scheme 
(thin lines with diamonds for the charm sector and triangles for bottom). 
Note, for case ii) the large factors used to scale the model calculations 
to fit in the plotting range.
The data are from the compilation published by the Particle Data Group (PDG) 
\cite{pdg}. The prompt J/$\psi$ measurement J/$\psi^{pr}$ is from the 
L3 experiment \cite{l3_qq}.}
\label{fig1}
\end{minipage} \end{tabular}
\end{figure}

In Fig.~\ref{fig1} we show a comparison of data \cite{pdg} and model prediction
for charmed and bottom hadron yields in e$^+$e$^-$ annihilations at 
$\sqrt{s}$=91 GeV.
For the model we have used the parameter set: $T$=170 MeV, $V$=16 fm$^3$ 
and $\gamma_s$=0.66, which represents the best fit of multiplicities
of hadrons with lighter quarks \cite{aa08_ee}.

We first note that the calculation employing the 5-flavor scheme is in 
very good agreement with the data, as demonstrated by the good $\chi^2$ 
per degree of freedom between the model and the data (excluding the $\Upsilon$ 
and prompt J/$\psi$) of 21.7/16 (34/18 when including all species)
(see also ref.~\cite{becattini08}).
Despite this overall agreement, the exceptions are significant: 
the $\Upsilon$ meson yield is underpredicted by the model by 17 orders 
of magnitude, while the prompt J/$\psi$ yield \cite{l3_qq} is underpredicted 
by almost 2 orders of magnitude.
Obviously, the production of quarkonia is expected to be strongly suppressed 
in the statistical model.
The disagreement is a consequence of the separate hadronization of the
$c$ and $\bar{c}$ quarks.
The measured prompt J/$\psi$ production in $Z^0$ decays (into hadrons) 
is about 3$\times10^{-4}$ \cite{l3_qq}.
The thermal model predicts a prompt yield for J/$\psi$ of 4.1$\times10^{-6}$
(1.6$\times10^{-7}$ for $\psi'$ and 4.3$\times10^{-7}$ for $\chi_{c1}$),
identically for the two calculation schemes.
Whenever the model seems to describe the yields of charmonia the measured 
yields are dominated by the feed down from bottom hadrons and the agreement 
only reflects the agreement seen for the open bottom hadrons and their 
branching ratios to charmonia, properly considered in the model.

The calculation employing a purely thermal ansatz underpredicts all
the measurements by many oders of magnitude, while for the light quark sector 
the differences between calculations with a pure thermal model and with 
the 5-flavor quark-antiquark scheme were found to be small \cite{aa08_ee}.
The strangeness suppression factor, which for the present results  
only enters in the calculation of the yields of $D_s$ and $B_s$ mesons, 
appears to have no counterpart in the heavy quark sector.
This reflects the fact that a negligible number of $c$ and $b$ quarks are
formed in the fragmentation process. In this case, the thermal weights 
describe the distribution of the initial quarks into heavy flavor hadrons.
Chemical equilibration is not required in this process.

For the case of hadron production in elementary hadronic collisions we employ 
the canonical realization of the thermal model \cite{gor,aa2,gra}.
For the description of the relative production cross sections of heavy flavored
hadrons, the energy dependence of the temperature parameter is the only model 
input, which is taken in a parametrized form from the fits of 
($u$,$d$,$s$)-hadron abundancies in central nucleus-nucleus 
collisions \cite{aat}.
For c.m. energies beyond 10 GeV per nucleon pair in nucleus-nucleus collisions 
a limiting temperature $T_{lim}$=164$\pm$5 MeV is reached. 
Recent fits of hadron yields in pp collisions \cite{kraus08} give very similar
values, independent of anergy. 

\begin{figure}[htb]
\begin{tabular}{cc} \begin{minipage}{.38\textwidth}
\hspace{-.5cm}\includegraphics[width=1.5\textwidth]{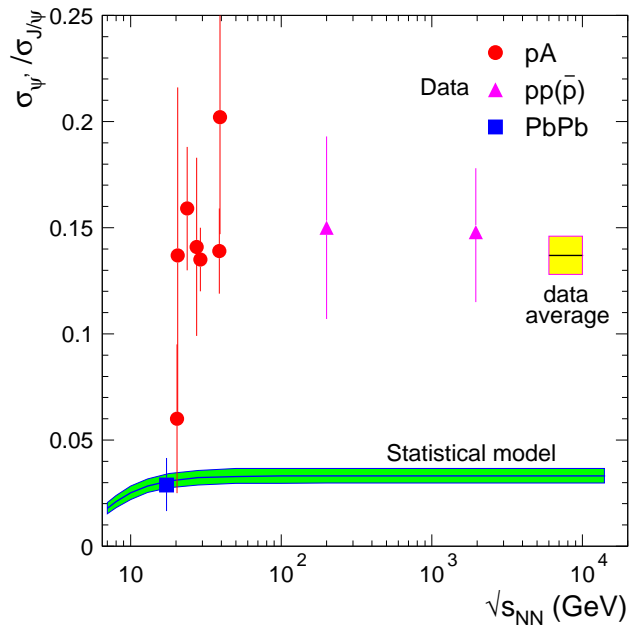}
\end{minipage}  & \begin{minipage}{.56\textwidth}
\vspace{-.9cm}
\caption{Production cross section of $\psi'$ relative to $J/\psi$.
The data for pA collisions are from the compilation by Maltoni et al. 
\cite{maltoni}; the points for elementary collisions are from the PHENIX 
experiment at RHIC \cite{phenix_psip} and from the CDF experiment at Tevatron
\cite{cdf_psip} (see text); the data point for Pb+Pb collisions at the SPS 
energy is from the NA50 experiment \cite{na50}.
The average value of the  pA and pp($\bar{\mathrm{p}}$) measurements with the 
corresponding error (see text) is represented by the shaded box.
The band denotes statistical model calculations for the 
temperature parametrization from heavy-ion fits \cite{aat} ($T_{lim}$=164 MeV) 
with $\pm$5 MeV error.}  
\label{fig2}
\end{minipage} \end{tabular}
\vspace{-.8cm}
\end{figure}

In Fig.~\ref{fig2} we show the model comparison to data for the relative 
production cross section of $\psi'$ and $J/\psi$ charmonia\footnote{The charm 
production cross section, which is an important model input parameter for 
the calculations of absolute yields, cancels out for this ratio.}.
The measurements in pA and pp($\bar{\mathrm{p}}$) collisions are 
above the model values by about a factor 4 
(corresponding to 10 experimental standard deviations; the average value 
of the measurements is 0.137$\pm$0.009, with a $\chi^2$ per degree of freedom 
of 0.88).
The relative production cross sections of charmonium states, as are
observed in all measurements in hadronic collisions cannot be described in 
the thermal approach. The temperature needed to explain the data would be
300 MeV, well above the Hagedorn limiting temperature, which is about 200 MeV.
This is in sharp contrast to the (only currently existing) measurement in 
central nucleus-nucleus collisions, performed at the SPS by the NA50 
experiment \cite{na50}, which is well described.
We note that the pA data exhibit a constant $\psi'/J/\psi$ production
ratio as a function of energy. 
In the model, the value is determined only by the temperature and this is 
reflected in the slight decrease of the ratio towards low energies. 
A constant value, also up to the LHC energies, is predicted
beyond $\sqrt{s_{NN}}\simeq$20 GeV.

 The measurements reported in Fig.~\ref{fig2} demonstrate that the relative 
production cross section $\psi'$/$J/\psi$ is identical in pA and in 
pp($\bar{\mathrm{p}}$) collisions, implying no visible influence of the cold 
nuclear medium.
Note that the ratio for the Tevatron energy was derived from the CDF 
measurements of $J/\psi$ \cite{cdf_jpsi} and $\psi'$ \cite{cdf_psip} and 
is for transverse momentum $p_t>$1.25 GeV/c (we have extrapolated the $\psi'$
measurement from 2 GeV/c down to 1.25 GeV/c).

\section{The statistical hadronization model in nucleus-nucleus collisions} 


The model has the following input parameters:
i) characteristics at chemical freeze-out: temperature, $T$, 
baryochemical potential, $\mu_b$, and volume corresponding to one unit 
of rapidity $V_{\Delta y=1}$, extracted from thermal fits of non-charmed 
hadrons \cite{aat};
ii) the charm production cross section in pp collisions, taken either 
from NLO pQCD calculations \cite{cac,rv1} or from experiment \cite{cc1},
used to calculate the number of directly produced $c\bar{c}$ pairs 
$N_{c\bar{c}}^{dir}$.
 The heavy quark (charm or bottom) balance equation \cite{pbm1} 
(including canonical suppression factors \cite{gor,aa2}, $I_1/I_0$):
\begin{equation}
N_{c\bar{c}}^{dir}=\frac{1}{2}g_c N_{oc}^{th}
\frac{I_1(g_cN_{oc}^{th})}{I_0(g_cN_{oc}^{th})} + g_c^2N_{c\bar c}^{th}.
\label{aa:eq1}
\end{equation}
is used to determine the fugacity factor $g_c$ ($I_n$ are modified Bessel 
functions; $N_{oc}^{th}$ and $N_{c\bar c}^{th}$ are the numbers of open and 
hidden charm hadrons, respectively, in the volume $V_{\Delta y=1}$, computed 
from their grand-canonical densities). 

\begin{figure}[htb]
\begin{tabular}{cc}
\begin{minipage}{.5\textwidth}
\hspace{-.8cm}\includegraphics[width=.98\textwidth]{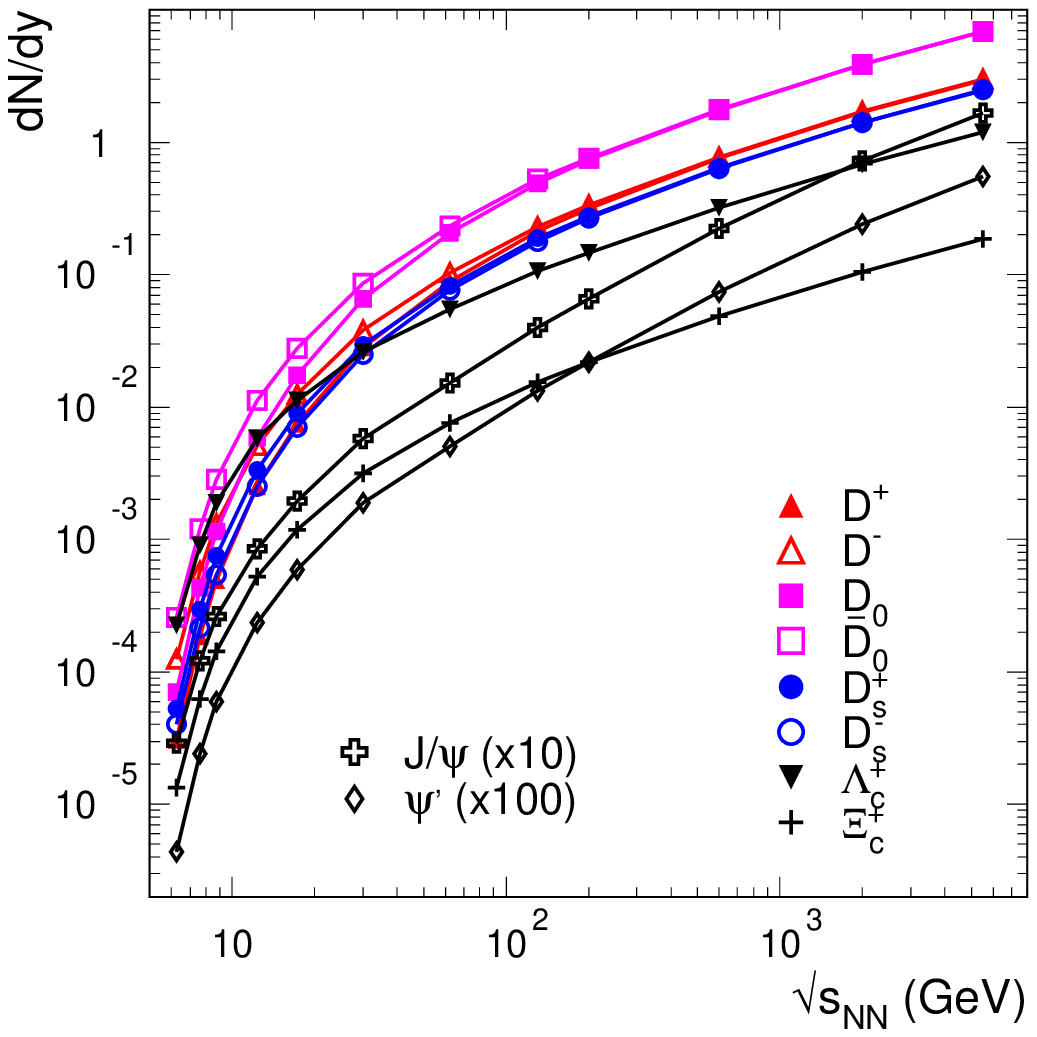}
\end{minipage} &\begin{minipage}{.5\textwidth}
\hspace{-.6cm}\includegraphics[width=1.0\textwidth]{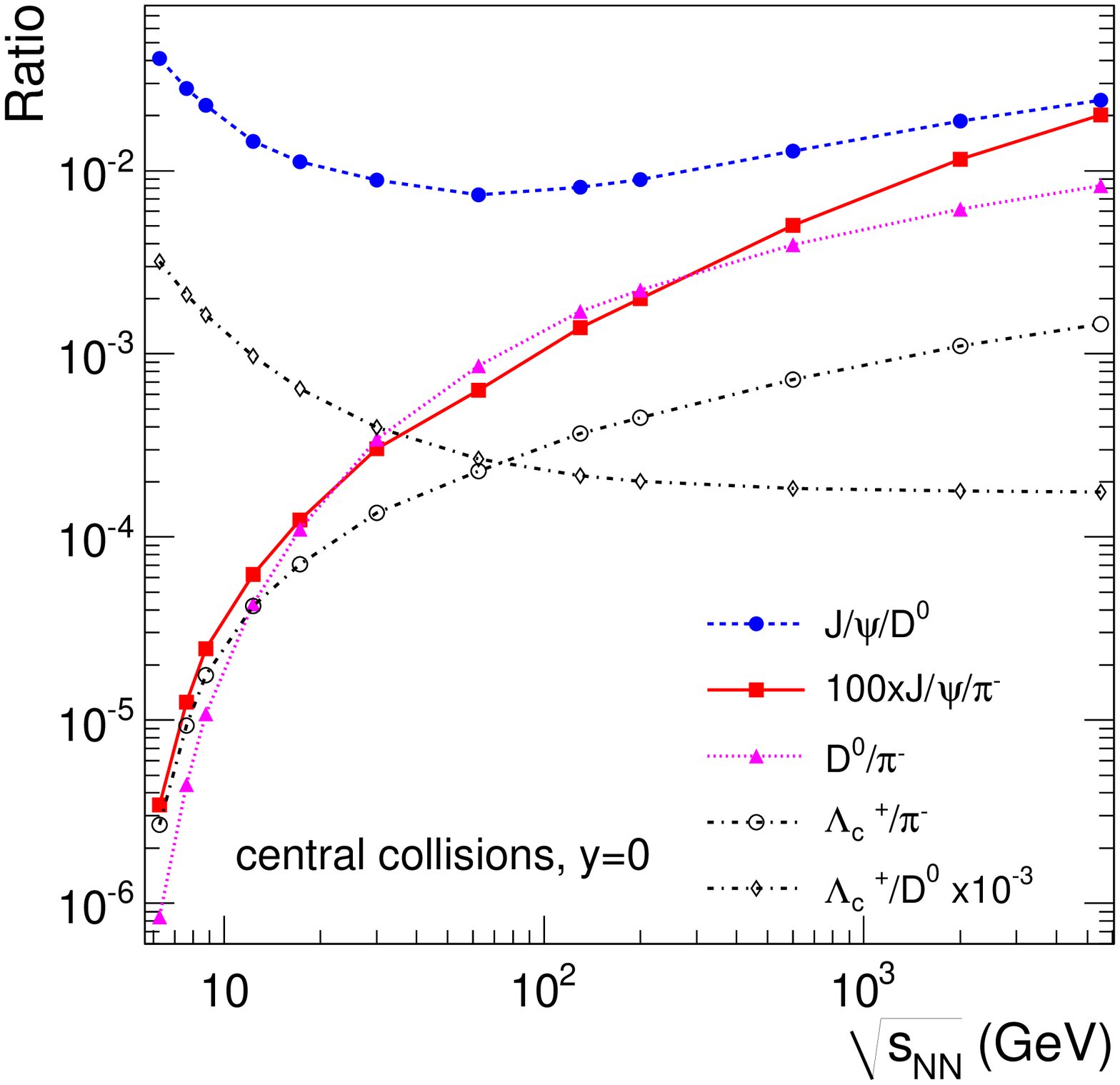}
\end{minipage}\end{tabular}
\caption{Left panel: energy dependence of the production yield at midrapidity
for open and hidden charm hadrons (with the energy dependence of 
${\mathrm d}\sigma_{c\bar{c}}/{\mathrm d} y$ as in \cite{aa2}).
Right panel: ratios calculated from midrapidity yields.}
\label{aa_fig3}
\end{figure}

A comprehensive set of model predictions as a function of energy, from the 
charm production threshold up to the LHC, is presented 
in Fig.~\ref{aa_fig3} for central Au-Au collisions ($N_{part}$=350).  
The left panel shows our predictions for the energy dependence of midrapidity 
yields for various open and hidden charm hadrons.
The most striking behavior is observed for the production of 
$\Lambda_c^+$ baryons: their yield relative to other charmed hadrons rises 
significantly towards lower energies.
In our approach this is caused by the increase in baryochemical potential 
towards lower energies (coupled with the charm neutrality condition). 
A similar behavior is seen for the $\Xi_c^+$ baryon.  
The relative production yields of D-mesons depend on their quark content 
and depend on energy only around threshold.
These results emphasize the importance of measuring, at low energies, 
in addition to D-mesons, also the yield of charmed baryons to get 
a complete measure of the total charm production cross section.

One of the motivations for the study of charm production at low energies
was the expectation \cite{cbm1,tol} to provide, by a measurement of D-meson 
production near threshold, information on their possible in-medium 
modification near the phase boundary. 
However, the cross section $\sigma_{c \bar c}$ is governed by the mass of 
the charm quark $m_c \approx 1.3$ GeV, which is much larger than any soft 
QCD scale such as $\Lambda_{QCD}$. 
Therefore we expect no medium effects on this quantity.
The much later formed D-mesons, or other charmed hadrons, may well change
their mass in the hot medium. 
Whatever the medium effects may be, they can, because of the charm 
conservation, $\sigma_{c \bar c} = \frac{1}{2} ( \sigma_D +
\sigma_{\Lambda_c} + ...) + ( \sigma_{\eta_c} +
\sigma_{J/\psi} + ...)$, expressed in Eq.~(\ref{aa:eq1}), 
in first order only lead to a redistribution of charm quarks \cite{aa2}.
In contrast, as a consequence of in-medium masses of open charm hadrons, 
the yields of charmonia can vary by up to 40-50\% and this is 
more prominent at threshold energies \cite{aa2}.

In the right panel of Fig.~\ref{aa_fig3} we show the energy dependence of
ratios of mid-rapidity yields of open and hidden charm hadrons and pions. 
The relative production of charmed hadrons to pions exhibits
a monotonic dependence on energy, unlike equivalent ratios involving strange
hadrons \cite{aat}. The different bahavior compared to the strangeness sector
is mainly due to the strong canonical suppression of open charm at low 
energies (determined in turn by strong decrease of the charm production cross 
section at low energies).
A non-monotonic behavior is observed in the production ratio of $J/\psi$ to
$D^0$, determined by the interplay of the strong canonical suppression of
open charm at low energies and the gradual onset as a function of energy
of the quadratic term in Eq.~(\ref{aa:eq1}).

\begin{figure}[htb]
\vspace{-.5cm}
\centering\includegraphics[width=.8\textwidth]{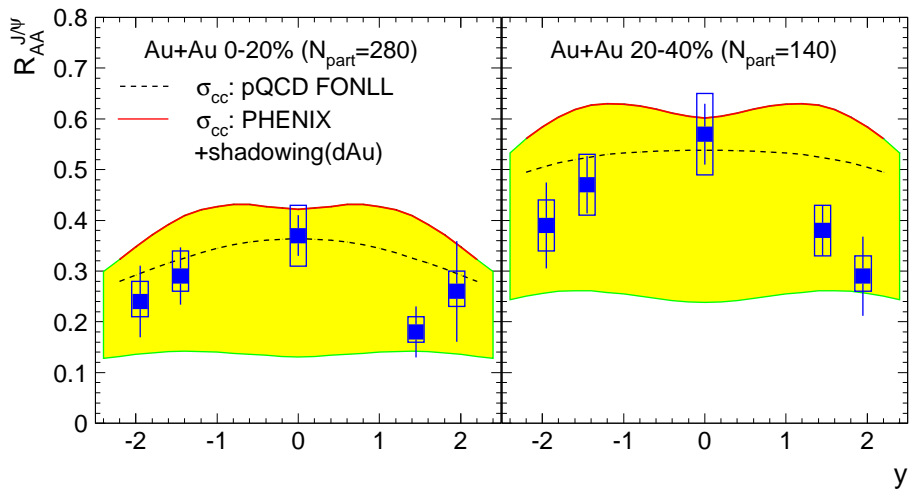}
\vspace{-.5cm}
  \caption{Rapidity dependence of $R_{AA}^{J/\psi}$ for two centrality 
classes. The data from the PHENIX experiment \cite{phe1} (symbols with errors)
are compared to calculations (lines, see text). The shaded area corresponds to 
calculations for the lower limit of the charm cross section as measured by 
PHENIX \cite{cc1}, with our shadowing scenario.}
\label{aa_fig1}
\end{figure}

In Fig.~\ref{aa_fig1} we present the rapidity dependence of the nuclear 
modification factor $R_{AA}^{J/\psi}$ (we use experimental data for the pp
reference).
While earlier \cite{aa2} we have compared data \cite{phe1} to our model 
predictions for the pQCD charm production cross section \cite{cac} 
(dashed line in Fig.~\ref{aa_fig1}), we use here as alternative the charm 
cross section as measured by PHENIX in pp collisions \cite{cc1} and consider 
in addition shadowing for Au-Au collsions as extracted from recent 
dAu data \cite{phe2}; we assume that the deviation of $R_{dAu}^{J/\psi}$ from 
unity is entirely due to shadowing.
Also in this case, our model describes the observed suppression and its 
rapidity dependence.  
The maximum of $R_{AA}^{J/\psi}$ at midrapidity is in our model due to the
enhanced generation of charmonium around mid-rapidity, determined by the
rapidity dependence of the charm production cross section. In this sense, 
the above result constitutes strong evidence for the statistical
generation of J/$\psi$ at chemical freeze-out.

\begin{figure}[htb]
\begin{tabular}{cc}
\begin{minipage}{.49\textwidth}
\hspace{-.5cm}\includegraphics[width=1.0\textwidth]{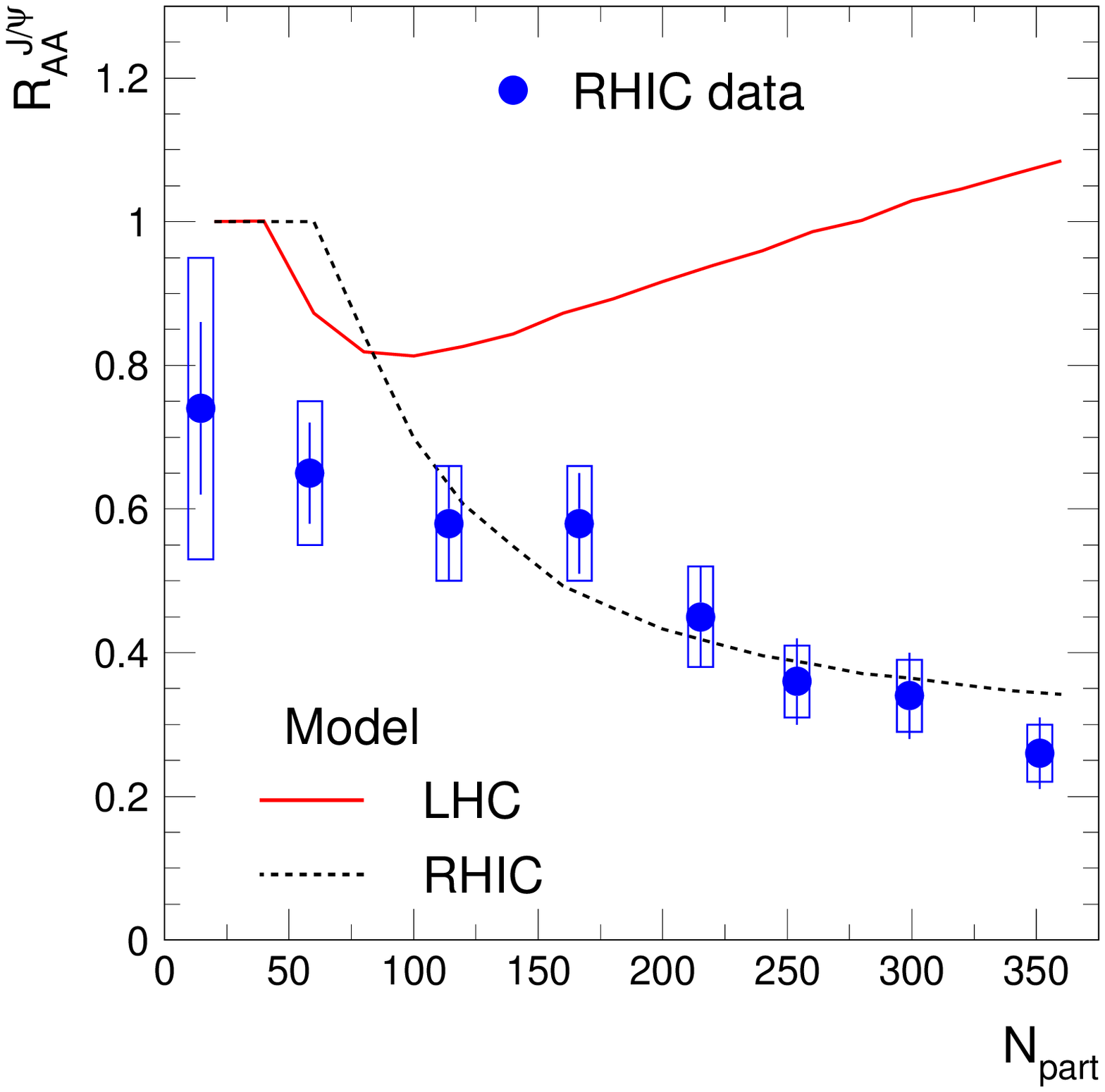}
\end{minipage}  & \begin{minipage}{.49\textwidth}
\hspace{-.5cm}\includegraphics[width=1.0\textwidth]{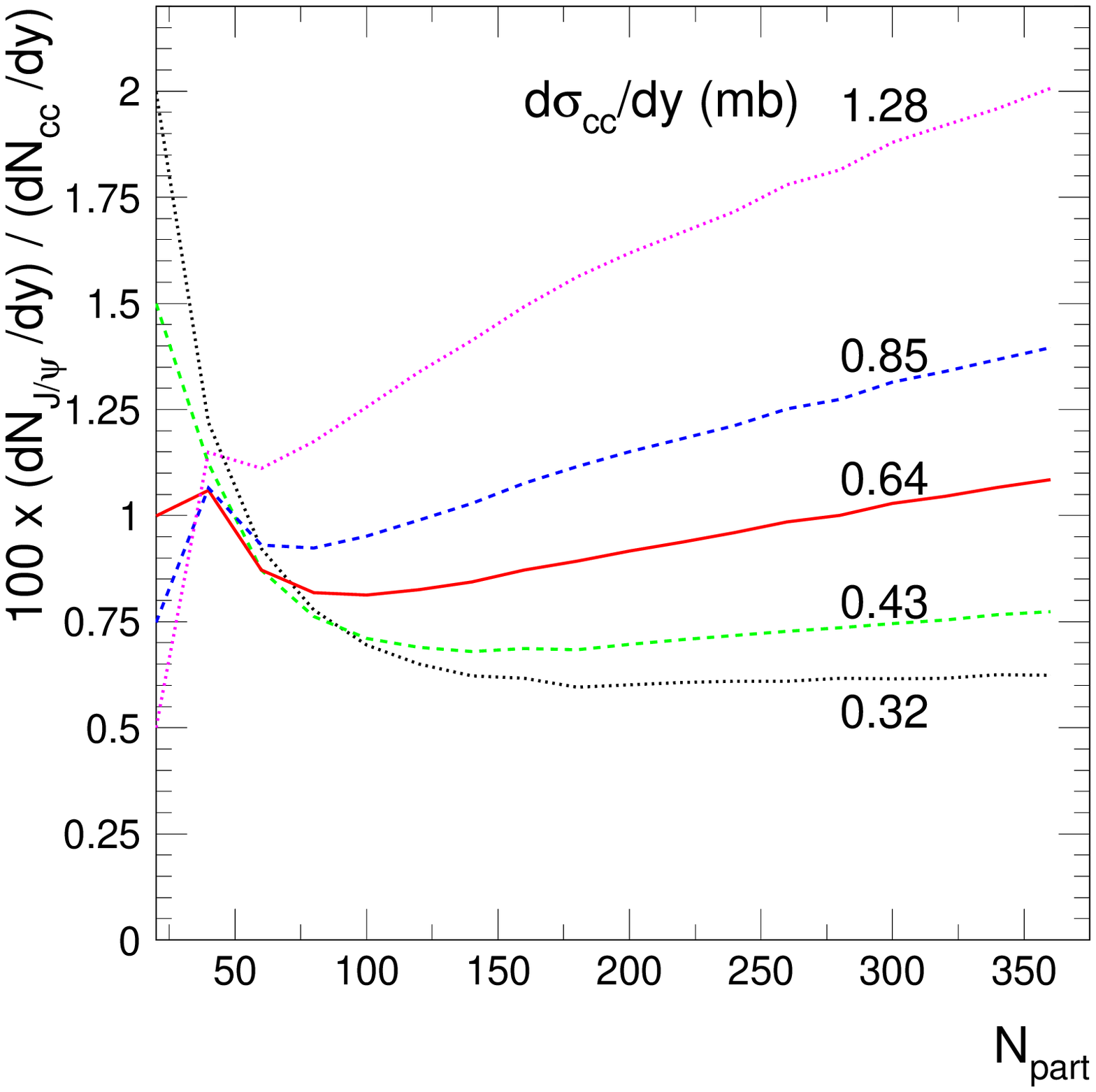}
\end{minipage}\end{tabular}
\vspace{-.5cm}
\caption{Centrality dependence of $R_{AA}^{J/\psi}$ for RHIC and LHC energies
(left panel) and of the $J/\psi$ rapidity density at LHC relative to the 
number of initially produced $c\bar{c}$ pairs (right panel, curves labelled 
by the $\ud\sigma_{c\bar{c}}/\ud y$) at midrapidity.}
\label{aa_fig2}
\end{figure}

The centrality dependence of $R_{AA}^{J/\psi}$ at midrapidity is shown in
the left panel of Fig.~\ref{aa_fig2}. Our calculations approach the value 
in pp collisions around $N_{part}$=50, which corresponds to an assumed minimal 
volume for the creation of QGP of 400 fm$^3$ \cite{aa2}.  
The model reproduces very well the decreasing trend versus centrality seen 
in the RHIC data \cite{phe1}.  
Note that, in our model, the centrality dependence of the nuclear modification 
factor arises entirely as a consequence of the still rather moderate rapidity 
density of initially produced charm quark pairs at RHIC 
($\ud N_{c\bar{c}}/\ud y$=1.6 for central collisions, using the FONLL charm 
production cross section \cite{cac}) leading to a marked centrality
dependence of the canonical suppression factor for open charm hadrons.
At the much higher LHC energy the charm production cross section (including 
shadowing in PbPb collisions \cite{rv1}) is expected to be about an order 
of magnitude larger.  
As a result, the opposite trend as a function of centrality is predicted, 
with $R_{AA}^{J/\psi}$ exceeding unity for central collisions.  A significantly 
larger enhancement of about a factor of 2 is obtained if the charm production 
cross section is two times larger than presently assumed, as seen in the right
panel of Fig.~\ref{aa_fig2}, where we show the $J/\psi$ production relative 
to the number of initially produced $c\bar{c}$ pairs.

\section{Conclusions}

We have shown that the statistical hadronization model describes well the 
measured decrease with centrality and the rapidity dependence of 
$R_{AA}^{J/\psi}$ at RHIC energy.
Importantly, well described in central heavy-ion collisions at SPS energy, 
where we have the only measurement to date,  is the ratio  $\psi'/J/\psi$.
In contrast, in elementary, e$^+$e$^-$ and in pp and p($\pi$)-nucleus, 
collisions the charmonium measurements cannot be explained within the 
statistical model.
This underlines the relevance of the statistical features in nucleus-nucleus
collisions.
Extrapolation to LHC energy leads, contrary to the observations at RHIC, 
to $R_{AA}^{J/\psi}$ increasing with collision centrality. If observed, such a
behavior will provide a dramatic proof of charmonium as the ultimate observable
to delineate the phase boundary of QCD matter.

\vspace{3mm}
K.R. acknowledges support from the Alexander von Humboldt Foundation.


\section*{References}
\begin{thebibliography}{99}
\bibitem{satz} T. Matsui, H. Satz, Phys. Lett. B {\bf 178} (1986) 416.

\bibitem{satz_kluberg} L. Kluberg, H. Satz, arXiv:0901.3831 [hep-ph].


\bibitem{rapp_rev} R. Rapp, D. Blaschke, P. Crochet, arXiv:0807.2470 [hep-ph].

\bibitem{aa2} A. Andronic, P. Braun-Munzinger, K. Redlich, J. Stachel, 
Nucl. Phys. A {\bf 789} (2007) 334; 
Phys. Lett. B {\bf 652} (2007) 259; 
Phys. Lett. B {\bf 659} (2008) 149. 

\bibitem{pbm_js_lb} P. Braun-Munzinger, J. Stachel, arXiv:0901.2500.

\bibitem{pbm1} P. Braun-Munzinger, J. Stachel,
Phys. Lett. B {\bf 490} (2000) 196; 
Nucl. Phys. A {\bf 690} (2001) 119c.

\bibitem{becattini97}
  F.~Becattini,
  arXiv:hep-ph/9701275;
  J.\ Phys.\ G {\bf 23} (1997) 1933. 

\bibitem{aa08_ee} A. Andronic, F. Beutler, P. Braun-Munzinger, K. Redlich, 
J. Stachel, Phys. Lett. B {\bf 675} (2009) 312. 

\bibitem{becattini08}
F. Becattini, P. Castorina, J. Manninen, H. Satz, Eur. Phys. J. C {\bf 56} 
(2008) 493. 

\bibitem{kraus08} I. Kraus, J. Cleymans, H. Oeschler, K. Redlich,
Phys. Rev. C {\bf 79} (2009) 014901. 

\bibitem{aa09} A. Andronic, F. Beutler, P. Braun-Munzinger, K. Redlich, 
J. Stachel, Phys. Lett. B {\bf 678} (2009) 350. 

\bibitem{fb09} F. Beutler, A. Andronic, P. Braun-Munzinger, K. Redlich, 
J. Stachel, arXiv:0910.1697.

\bibitem{pdg}
   C. Amsler {\it et al.}  (Particle Data Group),
  Phys. Lett. B {\bf 667} (2008) 1.

\bibitem{l3_qq} M. Acciarri et al. (L3 coll.), Phys. Lett. B {\bf 453} (1999) 94.

\bibitem{gor} M.I. Gorenstein, A.P. Kostyuk, H. St\"ocker, W. Greiner,
Phys. Lett. B {\bf 509} (2001) 277. 

\bibitem{gra} L. Grandchamp, R. Rapp, Phys. Lett. B {\bf 523} (2001) 60;
Nucl. Phys. A {\bf 709} (2002) 415. 

\bibitem{aat} A. Andronic, P. Braun-Munzinger, J. Stachel, 
Phys. Lett. B {\bf 673} (2009) 142; Erratum: ibid. 673 (2009) 516.

\bibitem{na50} B. Alessandro et al. (NA50 coll.), Eur. Phys. J. C {\bf 49} (2007) 559. 

\bibitem{maltoni} F. Maltoni et al., Phys. Lett. B {\bf 638} (2006) 202. 

\bibitem{phenix_psip} PHENIX collaboration, results presented at 
Quark Matter 2009 conference, {\tt http://www.phy.ornl.gov/QM09/}

\bibitem{cdf_jpsi} D. Acosta et al. (CDF coll.), Phys. Rev. D {\bf 71} (2005) 032001.

\bibitem{cdf_psip} T. Aaltonen et al. (CDF coll.), Phys. Rev. D {\bf 80} (2009) 031103.

\bibitem{cac} 
M. Cacciari, P. Nason, R. Vogt, Phys. Rev. Lett. {\bf 95} (2005) 
122001. 

\bibitem{rv1} ALICE Collaboration, J. Phys. G 32 (2006) 1295;
R. Vogt, Int. J. Mod. Phys. E {\bf 12} (2003) 211. 

\bibitem{cc1} A. Adare et al. (PHENIX coll.), Phys. Rev. Lett. {\bf 97} (2006) 252002.



\bibitem{cbm1} P. Senger, J. Phys. Conf. Series {\bf 50} (2006) 357.
\bibitem{tol} L. Tolos,  J. Schaffner-Bielich, H. St\"ocker,
Phys. Lett. B {\bf 635} (2006) 85. 

\bibitem{phe1} A. Adare et al. (PHENIX coll.),  Phys. Rev. Lett. {\bf 98} (2007) 232301. 
 
\bibitem{phe2} A. Adare et al. (PHENIX coll.), Phys. Rev. C {\bf 77} (2008) 024912; Erratum ibid. C {\bf 79} (2009) 059901 


\end{thebibliography}
\end{document}